\documentclass[%
 reprint,
 amsmath,amssymb,
 aps,
]{revtex4-2}
\UseRawInputEncoding
\usepackage{graphicx}% Include figure files
\usepackage{dcolumn}% Align table columns on decimal point
\usepackage{enumerate}
\usepackage{amsmath}
\usepackage{bm}% bold math
\begin{document}

\preprint{APS/123-QED}

\title{Cosmological implications of non-minimally coupled $f(Q)$ gravity}% Force line breaks with \\

\author{$^1$Santanu Das}
\author{$^2$Nilanjana Mahata}
\author{$^3$Priyanka Ray}
 \affiliation{$^1$Department of Basic science and Humanities,Institute of Engineering $\&$ Management,kolkata India and \\ Department of Mathematics,Jadavpur University,Kolkata,India.\\
 $^{2}$ Department of Mathematics,Jadavpur University,Kolkata,India.\\
 $^{3}$ Department of Mathematics,South Calcutta Girls' College,Kolkata,India.}
 %\affiliation{$^2$Department of Mathematics,Jadavpur University,Kolkata,India.}\\



\begin{abstract}
 We experience some challenges in general gravitational  theory owing to Einstein to explain late time acceleration of universe. To address this issue, geometric components of  gravity have been modified in quite a few occasions to have a more general structure with some freedom. One such approach is to change the geometric components of gravity where gravitational interaction is denoted by $Q$, $Q$ being the non metricity. In our work, we have considered symmetric teleparallel gravity i.e,  modified  the geometry with the help of non-metricity $Q$ or $f(Q)$ gravity. We have considered a specific form  $f(Q)$ which is nothing but the linear combination of $Q$ and $\alpha Q^n,n\neq 1$, where $Q$ is coupled with Lagrangian matter. Forming the autonomous system from governing equations and then solving it, we have  tried to analyze the nature of universe using dynamical system analysis. We have studied the behavior of the universe under several circumstances. Then, We have studied the stability around critical points  and considering the recent  observational data available for some cosmological parameters, feasible solutions are noted which depicts late time acceleration. We can see that $f(Q)$ gravity model can be considered as alternative model to $\Lambda$CDM model.
\end{abstract}

\maketitle

%\tableofcontents

\section{\label{sec:level I}Introduction:\protect}

Recent observational findings \cite{ag,pe} on late time acceleration of current universe\cite{sah} have posed  some theoretical  challenges to gravitational theories.  Although the general theory of relativity is the most successful theory of last century for the description of gravitational interaction, it  has been also struggling with several limitations like flatness, initial singularity etc. A satisfactory answer requires more generalized theory of gravity involving more general geometric structures. Thus, extended theories of gravity \cite{fs,sc} have been developed.\\  

Extended theories of gravitation \cite{khay} have been made through several modifications in Einstein's theory\cite{jwma,hiar} and in general these modified theories will have General Relativity as a particular case. These  theories are developed by modifying Einstein Hilbert action which yields Einstein filed equations and  these modifications  created interest amongst scientific community as scientists thought that it would be useful at scales close to Plank scale which is at very early universe near black hole singularity. But eventually, it provided a significant breakthrough in studying universe at low energies i.e, at large scale in late universe. Alternative gravitational theories are nothing but the attempts in introducing semi classical schemes in which most of the useful features of General Relativity can be identified.

Recent observational data depicts that approximate proportion of ordinary baryonic matter, dark matter and dark energy is respectively 4 \%, 20\% and 76\% \cite{as,sp,dj}. Dark matter has the clustering property like ordinary matter but dark energy is not describable  in that way. Both of them can be distinguished through energy conditions\cite{ld}. \\

In between two accelerated epoch there are decelerated expansion, radiation and matter dominated eras. Proportion of matter and energy in universe which are supported by observational data is astonishing.
Cosmological observations suggests that $\Lambda$CDM model is the best fitted model till date with some anomalies. To alleviate  those  anomalies in  $\Lambda$CDM model, other different cosmological models like Quintessence, K-essence, Coupled dark energy, Unified dark energy, f(R), f(R,T) etc models have been studied extensively i.e, to resolve  these issues, models have been developed in two approaches. In Einstein field equations, we are aware that there are two parts, one is matter part and another is gravity part. Some of the scientists have focused in modifying the matter part \cite{td,ni,gj,iz,cf,ss} and others were involved in modifying the gravity part \cite{cha,sn,bo,bra}. Thus modifications in gravity\cite{ens,amus,nm} is a different approach and  interesting way to explain universe's late time accelerated expansion whose limiting conditions can be obtained as General Relativity(GR).\\
 
We know that gravitational phenomena is represented by curved spacetime. Along with curvature, torsion and non metricity are associated with the connections of a metric space. In standard General Relativity which is  due to Einstein, both non metricity and torsion vanish and in  this framework, Ricci curvature R is the basic block of  spacetime.  Another approach is that where gravity is described by torsion T only. A third alternative representation is developed on a flat spacetime without curvature and torsion, where non metricity Q represents the gravitational interaction, known as the symmetric teleparallel gravity.  It is to be remembered that non metricity is the property of connection, not of the manifold. Jimenez et al.\cite{nonmet} have developed the symmetric teleparallel gravity into f(Q) gravity.\\

We have considered here cosmological model where gravitational interactions are represented by non-metricity $Q$ which has been taken into account for gravitational modifications. So, we are considering newly proposed symmetric  teleparallel  $f(Q)$ gravity\cite{is,la,ta,yu,nf,sm,ab,sana,wk,jm,ad,jw,kf,rfe,jl,san} where $Q$ is non-metricity \cite{jhk,lj,im}.\\

One key problem in the theory of gravity remains in finding the analytical or numerical solutions as there are lot of non linear terms in the field equations. That's why it is very difficult to find the analytical solutions of the field equations obtained from different cosmological models.
Dynamical system analysis\cite{lp} is such a tool \cite{ba,lo} which we can use to understand the qualitative nature of the physical system which is governed by the field equations which obtained from a specific cosmological model. \\In Dynamical system analysis\cite{jwai}, the field equations are transformed into first order autonomous system of ordinary differential equations. The stability conditions are analysed  around the critical points which are obtained from the system of aforesaid differential equations and then  by finding linearized jacobian matrix around those points and studying the respective eigen values. \\

Our motivation for this work is to incorporate late time cosmic acceleration for some functional forms of f(Q)  which  we have considered as a combination of $f_1(Q)$ and $f_2(Q)$, and analysing stability of the model. We all are aware that the equation of state parameter plays an important role in predicting different fluid description of universe. Similarly some other cosmological parameters play major roles in describing the evolution of universe. So here we shall be considering current observational data \cite{dm,gs,na,de,al,tmc,fda} of different cosmological parameters for analysing the model.\\
 
In this paper, we have discussed basic tools of $f(Q)$ gravity in Section-II, implementations of $f(Q)$ gravity in FRW spacetime in Section-III, formation of autonomous system and analysis of critical points in Section-IV, conclusion in Section-V.\\

\section{\label{sec:level II}Basic tools of $f(Q)$ gravity:\protect}

General theory of gravity is based on metric theory and in this case connection is symmetric and metric dependent. However, different approaches can be considered to characterize the space-time. One of such type of approaches is to consider symmetric teleparallel ($\int d^4 x \sqrt{-g} Q$)action instead of normal Einstein-Hilbert action. Here the modified gravity theory is dependent on nonlinear extension of  non-metric scalar $Q$ and the corresponding gravity theory is noted as $f(Q)$ gravity theory.\\
 In this work, we consider the action for matter coupling\cite{hf} in $f(Q)$ gravity with the help of two functions which is defined by \cite{th}
 \begin{equation}
 S=\int d^4 x \sqrt{-g} [\frac{1}{2}f_1(Q)+f_2(Q) L_M] 
  \end{equation}
  where $g$ is the determinant of the metric, $f_1(Q)$ and $f_2(Q)$ are two arbitrary functions of the non metricity $Q$, $L_M$ is the Lagrangian function for the matter field.\\
  Here the non-metricity tensor and the traces are 
   \begin{equation}
Q_{\alpha \beta \gamma}=\nabla_{\alpha}g_{\beta \gamma} 
  \end{equation}
    \begin{equation}
Q_{\alpha}=Q_{\alpha}^{\beta} \hspace{0.10cm} {_\beta},\Tilde{Q}_{\alpha}=Q^{\alpha}_{\alpha\beta}
  \end{equation}
  Superpotential can be introduced as a function of non metric tensor with the following equation
  \begin{equation}
4 P_{\beta \gamma}^{\alpha}=-Q_{\beta \gamma}^{\alpha}+2 Q_{(\beta^{\alpha} \gamma)}-Q^{\alpha}g_{\beta \gamma}-\tilde {Q}^{\alpha}g_{\beta \gamma}-\delta_{(\beta^{ \large{Q}_{\gamma})}}^{\alpha}
  \end{equation}
  where the trace of the nonmetricity tensor takes the form 
  \begin{equation}
  Q=-Q_{\alpha \beta \gamma}P^{\alpha \beta \gamma}
  \end{equation}
  For simplicity, let us consider the following equations 
   \begin{equation}
 f=f_{1}(Q)+2f_{2}(Q)L_{M}
  \end{equation}
   \begin{equation}
 F=f^{'}_{1}(Q)+2f^{'}_{2}(Q)L_{M}
  \end{equation}
  where ($'$) signifies the derivatives of the functions
  $f_{1}(Q)$ and $f_{2}(Q)$ with respect to $Q$.\\To identify the fluid description of the space-time, energy-momentum tensor can be introduced as 
  \begin{equation}
  T_{\beta \gamma}=-\frac{2}{\sqrt{-g}}\frac{\delta{(\sqrt{-g}L_{M})}}{\delta g^{\beta \gamma}}
  \end{equation}  
  By varying the action, considered in equation (1) with respect to the metric tensor, following gravitational field equation can be obtained
  \begin{equation}\label{eq9}
  \begin{split}
  \frac{2}{\sqrt{-g}}\nabla_{\alpha}(\sqrt{-g}FP^{\alpha}_{\hspace{0.07cm}\beta\gamma})+\frac{1}{2}g_{\beta \gamma}f_{1} &\\ +F(P_{\beta \mu\nu}Q_{\gamma}^{\hspace{0.2cm}\mu\nu}-2Q_{ \mu\nu\beta}P^{\mu\nu}_{\hspace{0.2cm}\gamma})=-f_{2}T_{\beta\gamma}
  \end{split}
  \end{equation}
  Now, equation (9) can be used for different cosmological applications in $f(Q)$ modified gravity.
  \section{\label{sec:level III}Implementation  of $f(Q)$ gravity in FRW space-time:\protect}
  For exploring several cosmological implementation,we consider the isotropic, homogeneous, spatially flat line element as 
  \begin{equation}
  ds^{2}=-N^{2}(t)dt^{2}+a^{2}(t)\delta_{ij}dx^{i}dx^{j}
  \end{equation}
  Here $N(t)$ denotes the lapse function and in the present case due to usual time reparametrization freedom, we may impose $N=1$ in any time. $\delta_{ij}$ denotes Kronecker delta and i, j run over the spatial components.\\
  Now, it is customary to define expansion and dilation parameters as
   \begin{equation}
  H=\frac{\dot{a}}{a} ~~and ~~T=\frac{\dot{N}}{N}
\end{equation} 
In this current line element, the non-metricity $Q$ is obtained  as
   \begin{equation}
  Q=6H^2
\end{equation}
Here, we shall be considering standard perfect fluid matter whose energy-momentum tensor, given by (8) is diagonal. So, the gravitational field equation (9) will imply here  two generalized Friedman equations 
obtained as
   \begin{equation}
f_{2}\rho=\frac{f_{1}}{2}-6F \frac{H^2}{N^2}
\end{equation}

   \begin{equation}
-f_{2} p=\frac{f_{1}}{2}-\frac{2}{N^2}[(\dot{F}-FT)H+F(\dot{H}+3H^2)]
\end{equation}
Here $\rho$ is the energy density and $p$ is the pressure of the fluid content in the space-time. It is very trivial to verify that by putting $f_{1}=-Q$ and $f_2=1=-F$, the aforesaid Friedman equations (13) and (14) reduce to the standard Friedman equations.\\
 The continuity equation for motion for matter field  can be derived from (13) and (14) as 
 \begin{equation}
 \dot{\rho}+3H(\rho+p)=-\frac{6f_{2}^{'}H}{f_{2}N^2}(\dot{H}-HT)(L_M+\rho)
 \end{equation}
From (15), the standard continuity equation can be derived by substituting $L_{M}=-\rho$ as
\begin{equation}
\dot{\rho}+3H(\rho +p)=0
\end{equation}
This is now compatible in accordance with the homogeneous and isotropic nature of the universe.\\
Here, we shall be proceeding with $N=1$. As, we are working in the framework of standard Friedman-Robertson-walker (FRW) geometry, dilation parameter $T$ and the non-metricity $Q$ are transformed to \begin{equation}
T=0 ~and ~  Q=6H^2 
\end{equation} respectively.
Hence the equations (13) and (14) can be reframed as 
\begin{equation}
3H^2=\frac{f_2}{2F}(-\rho+\frac{f_1}{2f_2})
\end{equation}
\begin{equation}
\dot{H}+3H^2+\frac{\dot{F}}{F}H=\frac{f_2}{2F}(p+\frac{f_1}{2f_2})
\end{equation}
 \section{\label{sec:level IV}Formation of autonomous system and stability analysis by Dynamical system approach:\protect}
 In this section, we confine ourselves to the dynamical system approach to analyse the stability of the system using linearized Jacobian matrix owing to Jacobi. We consider two arbitrary functions of the non metricity $Q$ as 
 \begin{equation}
 f_1(Q)=\alpha Q^n,n\neq 1 ~ and~ f_2(Q)=Q
 \end{equation}
 Here $\alpha$ and $n$ are arbitrary constants.\\
 
Let us introduce  following dimensionless variables
\begin{equation}
 x=-\frac{-\rho f_2}{QF} ~and~ y=\frac{f_1}{2QF}
 \end{equation}
 \subsection{Autonomous system with equation of state parameter}
Let us also consider the equation of state parameter $p=\rho ~\omega$ where $\rho$ is the energy density and $p$ is the pressure of fluid content in the space-time.
Using equations (21) and (17), in the gravitational field equations (18) and (19) and continuity equation (16), we can formulate following system of autonomous equations
\begin{equation}
x^{'}=-\frac{11}{3}\epsilon x -3x\omega-3xy+3\omega x^2
\end{equation}
\begin{equation}
y^{'}=-y\epsilon (2n-1)-3y^2+3xy\omega+3y
\end{equation}
Here($'$) denotes the derivative with respect to $\eta=\ln a$ and
\begin{equation}
\epsilon=-\frac{\dot{H}}{H^2}=q+1
\end{equation}

where $q$ is the deceleration parameter.\\
 At the very beginning, by introducing dimensionless variables, introduced in equation (21), we have obtained  autonomous system of equations in (22) and (23). From the aforesaid system we have obtained three set of critical points(in Table-I) through solutions of 
 \begin{equation}
 x^{'}=0 ,
 y^{'}=0
 \end{equation}
 Apart from the first point $(0,0)$, other two points are denoted by deceleration parameter, $q$(in equation (24), $n$ and equation of state parameter, $\omega$. By varying the values of these parameters, we have considered different cases for the aforesaid critical points and analyzed the stability of the universe in the current late time acceleration epoch. We have formed the linearized matrix and obtained the corresponding eigen values with respect to the critical points. Then considering different values of the parameters supported by current observational data, we have done the stability analysis with respect to the critical points locally.  \\
Solving the above autonomous problem by framing a linearized Jacobian matrix
\[    
J=
\begin{pmatrix}
    \frac{-11}{3}\epsilon-3\omega-3y+6\omega x & -3x \\
    3y\omega &  -\epsilon(2n-1)-6y+3x\omega+3  \\
    
\end{pmatrix}
\]

for three hyperbolic critical points, eigen values corresponding to those critical points are found which are represented in  table-I.
\begin{table}[htbp]
		\resizebox{\linewidth}{!}{
		\begin{tabular}{|c|c|}
		\hline

			\textbf{Critical points$(x,y)$} & \textbf{Eigen values} \\ \hline
			
			$( 0 , 0)$ &
			$3-\epsilon(2n-1),-\frac{11}{3}\epsilon-3\omega$\\ 
			\hline
			
			( $\frac{1.22\epsilon+\omega}{\omega} , 0)$ &
			$\frac{11}{3}\epsilon+3\omega,\frac{11}{3}\epsilon+3\omega-\epsilon(2n-1)+3$\\ 
			\hline
	$( 0 , 0.33\epsilon-0.67\epsilon n +1)$ &
			$2.01\epsilon n-3\omega-4.66\epsilon-3,4\epsilon n-2\epsilon-\epsilon (2n-1)-3$\\	
			\hline	
			
		\end{tabular}
		}
	
	\caption{Set of critical points and corresponding eigen values.}

\end{table}

Here critical points and  the corresponding eigen values depend on the predefined parameters $\epsilon=q+1$, $\omega$ and $n$.\\
Now, we shall be doing explicit stability analysis around those critical points. Critical points will vary with the value of above parameters. We have taken $q$ in the range of $-0.48$ to $-0.55$ as prescribed in \cite{na,ab,vm}. For varying $n$, we can find a form of $f(Q)$ for stable universe.\\
Critical point $(0,0)$ and the eigen values corresponding to $q$ and $n$ are given in table-II.

	\begin{table}[h!]

	\centering	
		\begin{tabular}{|c|c|c|c|c|c|}
			\hline
		\textbf{No}& \textbf{$q$} & \textbf{$\epsilon$} & \textbf{n} & \textbf{point} & \textbf{Eigen Value}\\ \hline
			
	$P_1$ & -0.53 & 0.47 & 2 & (0,0) & 1.59 ,-1.72-3$\omega$\\ \hline
	$P_2$ & -0.53 & 0.47 & 3 & (0,0) & 0.65 ,-1.72-3$\omega$\\ \hline	
	$P_3$ & -0.53 & 0.47 & 4 & (0,0) & -0.29 ,-1.72-3$\omega$\\ \hline
	$P_4$ & -0.53 & 0.47 & 5 & (0,0) & -1.23 ,-1.72-3$\omega$\\ \hline
	$P_5$ & -0.53 & 0.47 & 6 & (0,0) & -2.17 ,-1.72-3$\omega$\\ \hline
	$P_6$ & -0.48 & 0.52 & 2 & (0,0) & 1.44,-1.90-3$\omega$\\ \hline
	$P_7$ & -0.48 & 0.52 & 3 & (0,0) & 0.4,-1.90-3$\omega$
	\\  \hline
	$P_8$ & -0.48 & 0.52 & 4 & (0,0) & -0.64,-1.90-3$\omega$\\ \hline
		$P_9$ & -0.48 & 0.52 & 5 & (0,0) & -1.68,-1.90-3$\omega$\\  \hline	
	$P_{10}$ & -0.48 & 0.52 & 6 & (0,0) & -2.72,-1.90-3$\omega$\\  \hline

			\end{tabular}
			
			\caption{For $q=-0.53$, $q=-0.48$ and for n=$2,3,4,5,6 $, this table shows eigen values corresponding to the stationary point $(0,0)$. Here the eigen values depend on the value of equation of state $\omega$.}

			\end{table}

			According to Table-II, we can clearly observe that \\
			For $q=-0.53$:\\
			\begin{enumerate}[I.]
			\item For $n=2 ~and ~ 3$($P_1,P_2$ from Table-II), the point $(0,0)$ is saddle when $\omega>-0.57$ and unstable whenever $\omega<-0.57$.
			
			\item For $n\geq 4$($P_3,P_4,P_5$ from Table-II), $(0,0)$ will be stable node when $\omega>-0.57$ and saddle node when $\omega<-0.57$. 
			
			 \end{enumerate}
			 	\begin{figure}[h!]
\centerline{\includegraphics[scale=0.75]{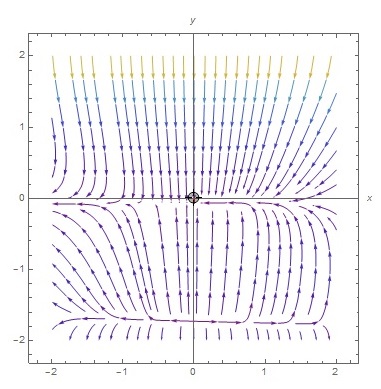}}
\caption{Phase plot corresponding to the point $(0,0)$ for $q=-0.53, n=20, \omega=-0.4$ which shows that for the choices of aforesaid values of $q, \omega~and~n$, $(0,0)$ is locally stable node}
\end{figure}
			 
			Similarly, for $q=-0.48$:\\	
			\begin{enumerate}[I.]
			\item For $n=2 ~and ~ 3$($P_6,P_7$ from Table-II), the point $(0,0)$ is saddle when $\omega>-0.63$ and  unstable whenever $\omega<-0.63$.
			
			\item For $n\geq 4$($P_8,P_9,P_{10}$from Table-II), $(0,0)$ will be stable node when $\omega>-0.63$ and saddle node when $\omega<-0.63$. 
		
			\end{enumerate}
\begin{figure}[h!]
\centerline{\includegraphics[scale=0.7]{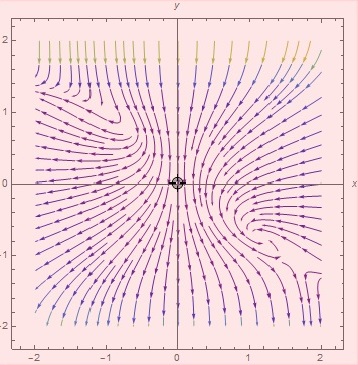}}
\caption{Phase plot corresponding to the point $(0,0)$ for $q=-0.48, n=3, \omega=-0.87$ ($P_{6}$ in Table-II) which shows that for  the choices of aforesaid values of $q, \omega~and~n$, $(0,0)$ is locally unstable node.}
\end{figure}
Critical point $(\frac{1.22\epsilon+\omega}{\omega},0)$ and the eigen values corresponding to $q$ and $n$ are given in table-III.
			\begin{table}
			[h!]
		
	\centering	
		\begin{tabular}{|c|c|c|c|c|c|}
			\hline
		\textbf{No}& \textbf{$q$} & \textbf{$\epsilon$} & \textbf{n} & \textbf{point} & \textbf{Eigen Value}\\ \hline
			$P_{11}$ & -0.53 & 0.47 & 2 & ($\frac{0.9044+\omega}{\omega}$,0) & 1.72+3$\omega$,3.3133+3$\omega$\\ \hline
			$P_{12}$ & -0.53 & 0.47 & 3 & ($\frac{0.9044+\omega}{\omega}$,0) & 1.72+3$\omega$,2.3733+3$\omega$\\ \hline	
			$P_{13}$ & -0.53 & 0.47 & 4 & ($\frac{0.9044+\omega}{\omega}$,0) & 1.72+3$\omega$,1.4333+3$\omega$\\ \hline	
			$P_{14}$ & -0.53 & 0.47 & 5 & ($\frac{0.9044+\omega}{\omega}$,0) & 1.72+3$\omega$,0.4933+3$\omega$\\ \hline	
				$P_{15}$ & -0.53 & 0.47 & 6 & ($\frac{0.9044+\omega}{\omega}$,0) & 1.72+3$\omega$,-0.4467+3$\omega$\\ \hline
		$P_{16}$ & -0.48 & 0.52 & 2 & ($\frac{0.6355+\omega}{\omega}$,0) & 1.91+3$\omega$,3.3467+3$\omega$\\ \hline	
	
	$P_{17}$ & -0.48 & 0.52 & 3 & ($\frac{0.6355+\omega}{\omega}$,0) & 1.91+3$\omega$,2.3067+3$\omega$\\ \hline	
		
		$P_{18}$ & -0.48 & 0.52 & 4 & ($\frac{0.6355+\omega}{\omega}$,0) & 1.91+3$\omega$,1.2667+3$\omega$\\ \hline
		
		$P_{19}$ & -0.48 & 0.52 & 5 & ($\frac{0.6355+\omega}{\omega}$,0) & 1.91+3$\omega$,0.2267+3$\omega$\\ \hline
		
	$P_{20}$ & -0.48 & 0.52 & 6 & ($\frac{0.6355+\omega}{\omega}$,0) & 1.91+3$\omega$,-0.8133+3$\omega$\\ \hline

			\end{tabular}
			
			\caption{For $q=-0.53$ , $q=-0.48$ and for n=$2,3,4,5,6 $, this table shows eigen values corresponding to the stationary point $(\frac{1.22\epsilon+\omega}{\omega},0)$. Here the eigen values depend on the value of equation of state parameter $\omega$.}
			\end{table}\

			According to Table-III,we can clearly observe that \\
			For $q=-0.53$:
			\begin{enumerate}[I.]
			\item For $n=2$ ($P_{11}$ Table-III), the point $(\frac{1.22\epsilon+\omega}{\omega},0)$ is stable when $\omega<-1.10$,saddle node when $-1.10<\omega<-0.57$ and unstable when $\omega>-0.57$. 
			\item For $n=3$($P_{12}$ in Table-III), the point $(\frac{1.22\epsilon+\omega}{\omega},0)$ becomes stable  when $\omega<-0.79$,saddle node when $-0.79<\omega<-0.57$ and unstable when $\omega>-0.57$. 
			\item For $n\geq 4$($P_{13},P_{14},P_{15}$ in Table-III), $(\frac{1.22\epsilon+\omega}{\omega},0)$ will be a stable node when $ \omega<-0.57$.
	  \item For $n=4$($P_{13}$ in Table-III), $(\frac{1.22\epsilon+\omega}{\omega},0)$ becomes an unstable node when $\omega>-0.48$ and saddle node when $-0.57<\omega<-0.48$.
		\item For $n=5$ ($P_{14}$ in Table-III), $(\frac{1.22\epsilon+\omega}{\omega},0)$ will be an unstable node when $\omega>-0.16$ and saddle node when $-0.57<\omega<-0.16$.
		\item For $n=6$($P_{15}$ in Table-III), $(\frac{1.22\epsilon+\omega}{\omega},0)$ turns to be an unstable node when $\omega>0.14$ and saddle node when $-0.57<\omega<0.14$
	
			\end{enumerate}
			
				\begin{figure}[h!]
\includegraphics[scale=0.7]{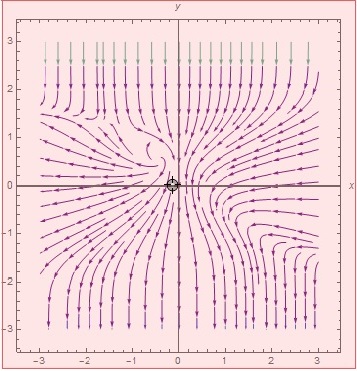}
\caption{Phase plot corresponding to the point $(-0.1305,0) $($P_{11}$ in Table-II)for $q=-0.53, n=2,\omega=-0.8$ which shows that for the choices of aforesaid values of $q, \omega~and~n$, $(-0.1305,0)$ is a saddle node.}
\end{figure}

			Similarly, $q=-0.48$:
			\begin{enumerate}[I.]
			
			\item For $n=2$ ($P_{16}$ Table-III), the point $(\frac{1.22\epsilon+\omega}{\omega},0)$ is stable when $\omega<-1.11$,saddle node when $-1.11<\omega<-0.64$ and unstable when $\omega>-0.64$. 
			\item For $n=3$($P_{17}$ in Table-III), the point $(\frac{1.22\epsilon+\omega}{\omega},0)$ becomes stable when $\omega<-0.77$,saddle node when $-0.77<\omega<-0.64$ and unstable when $\omega>-0.64$. 
			\item For $n\geq 4$($P_{18},P_{19},P_{20}$ in Table-III),$(\frac{1.22\epsilon+\omega}{\omega},0)$ will be a stable node when $ \omega<-0.63$.
	  \item For $n=4$($P_{18}$ in Table-III), $(\frac{1.22\epsilon+\omega}{\omega},0)$ becomes an unstable node when $\omega>-0.42$ and saddle node when $-0.63<\omega<-0.42$.
		\item For $n=5$ ($P_{19}$ in Table-III), $(\frac{1.22\epsilon+\omega}{\omega},0)$ will be an unstable node when $\omega>-0.07$ and saddle node when $-0.63<\omega<-0.07$.
		\item For $n=6$($P_{20}$ in Table-III), $(\frac{1.22\epsilon+\omega}{\omega},0)$ turns into an unstable node when $\omega>0.27$ and saddle node when $-0.63<\omega<0.27$.
			\end{enumerate}
			
	\begin{figure}[h!]
\includegraphics[scale=0.75]{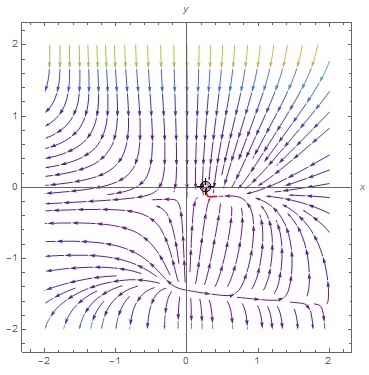}
\caption{Phase plot corresponding to the point $(0.2695,0)$ for $q=-0.48,n=10,\omega=-0.87$ which shows that for  the choices of aforesaid values of $q,\omega~and~n$, $(0.2056,0)$ is locally stable node.}
\end{figure}
	Critical point $(0,0.33\epsilon-0.67\epsilon n +1)$ and the eigen values corresponding to $q$ and $n$ are given in table-IV.		
			\begin{table}[h!]

	\centering	
		\begin{tabular}{|c|c|c|c|c|c|}
			\hline
		\textbf{No}& \textbf{$q$} & \textbf{$\epsilon$} & \textbf{n} & \textbf{point} & \textbf{Eigen Value}\\ \hline
		$P_{21}$ & -0.53 & 0.47 & 2 & (0,0.5253) & -3.3008-3$\omega$,-1.59\\ \hline
				
		$P_{22}$ & -0.53 & 0.47 & 3 & (0,0.2104) & -2.3561-3$\omega$,-0.65\\ \hline
		$P_{23}$ & -0.53 & 0.47 & 4 & (0,-0.1045) & -1.4114-3$\omega$,0.29\\ \hline
		$P_{24}$ & -0.53 & 0.47 & 5 & (0,-0.4194) & -0.4667-3$\omega$,1.23\\ \hline
			$P_{25}$ & -0.48 & 0.52 & 2 & (0,0.4748) & -3.3328-3$\omega$,-1.44\\ \hline				
		
		$P_{26}$ & -0.48 & 0.52 & 3 & (0,0.1264) & -2.2876-3$\omega$,-0.4\\ \hline
		$P_{27}$ & -0.48 & 0.52 & 4 & (0,-0.2220) & -1.2424-3$\omega$,0.64\\ \hline	
		
		$P_{28}$ & -0.48 & 0.52 & 5 & (0,-0.5704) & -0.1972-3$\omega$,1.68\\ \hline

			\end{tabular}
			
			\caption{For $q=-0.53$ , $q=-0.48$ and for n=$2,3,4,5,6 $,this table shows eigen values corresponding to the stationary point $( 0 , 0.33\epsilon-0.67\epsilon n +1)$. Here the eigen values depend on the value of equation of state $\omega$.}
			\end{table}
	According to Table-IV, we can clearly observe that for $q=-0.53$:\\
\begin{enumerate}[I.]	
	\item  For $n=2$($P_{21}$ in Table-IV), the point $(0,0.33\epsilon-0.67\epsilon n +1)$ is stable when $ \omega>-1.1$ and saddle node when $ \omega<-1.1$.
	\item For $n=3$($P_{22}$ in Table-IV), the point $(0,0.33\epsilon-0.67\epsilon n +1)$ will be stable  when $\omega>-0.78$ and saddle node when $\omega<-0.78$.
	\item For $n=4$($P_{23}$ in Table-IV),$(0,0.33\epsilon-0.67\epsilon n +1)$ becomes a saddle node for  $\omega>-0.47$ and an unstable node when $\omega<-0.47$.
	\item For $n=5$($P_{24}$ in Table-IV),$(0,0.33\epsilon-0.67\epsilon n +1)$ will be a saddle node when $\omega>-0.15$ and an unstable node when $\omega<-0.15$.
	
	\end{enumerate}
	\begin{figure}[h!]
\includegraphics[scale=0.7]{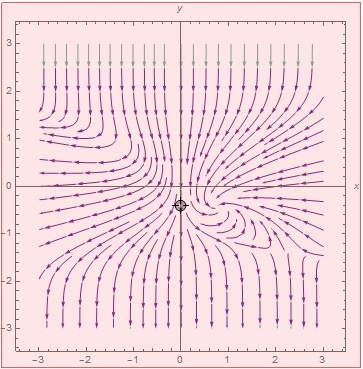}
\caption{Phase plot corresponding to the point $(0,-0.4194)$($P_{24}$ in Table-IV) for $q=-0.53,n=5,\omega=-0.7$ which shows that for  the choices of aforesaid values of $q,\omega~and~n$, $(0,-0.4194)$ is an unstable node.}
\end{figure}
			Similarly, $q=-0.48$:\\
\begin{enumerate}[I.]	

\item  For $n=2$($P_{25}$ in Table-IV),the point $(0,0.33\epsilon-0.67\epsilon n +1)$ is stable when $ \omega>-1.11$ and saddle node when $ \omega<-1.11$.
	\item For $n=3$($P_{26}$ in Table-IV), the point $(0,0.33\epsilon-0.67\epsilon n +1)$ becomes stable  when $\omega>-0.76$ and saddle node when $\omega<-0.76$.
	\item For $n=4$($P_{27}$ in Table-IV),$(0,0.33\epsilon-0.67\epsilon n +1)$ turns to be a saddle node for $\omega>-0.41$ and an unstable node when $\omega<-0.41$
	\item For $n=5$ ($P_{28}$ in Table-IV),$(0,0.33\epsilon-0.67\epsilon n +1)$ will be a saddle node when $\omega>-0.06$ and an unstable node when $\omega<-0.06$.

			\end{enumerate}
			\begin{figure}[h!]
\includegraphics[scale=0.7]{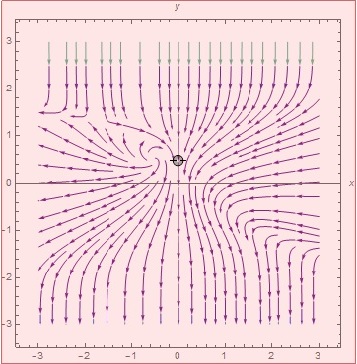}
\caption{Phase plot corresponding to the point $(0,0.4748)$($P_{25}$ in Table-IV) for $q=-0.48,n=2,\omega=-0.75$ which shows that for the choices of aforesaid values of $q,\omega~and~n$, $(0,0.4748)$ is locally stable  node.}
\end{figure}
\subsection{Autonomous system with equation of state parameter in terms of  dimensionless variables}
Here,equation of state parameter
\begin{equation}
\omega=\frac{p}{\rho}=\frac{-2\dot{H}-3H^2}{3H^2}
\end{equation}
Using (18) and (19) in (25), we obtain
\begin{equation}
\omega(1+\frac{f_{2}\rho}{3H^2F})=-\frac{f_{2}\rho}{3H^2F}+\frac{2\dot{F}}{3FH}-1
\end{equation}
With the help of the choices of $f_1(Q)$ and $f_2(Q)$ in (20) and dimensionless variables in (21),(26) yields the expression for equation of state parameter $\omega$ as 
\begin{equation}
\omega=\frac{-2x-\frac{8}{3}(n-1)^{2}\epsilon y+1}{2x-1}
\end{equation}
where $\epsilon$ is defined in (24).
Substituting (27) in (22) and (23), autonomous system for this current model transformed into 
\begin{align}
x^{'}=-\frac{11}{3}\epsilon x-3xy+\frac{9x^2}{2x-1}+\frac{8(n-1)^2 \epsilon y x}{2x-1}-\frac{3x}{2x-1}
 \notag\\ -\frac{6x^3}{2x-1}-\frac{8(n-1)^2 \epsilon y x^2}{2x-1}
\end{align}

\begin{align}
y^{'}=-y\epsilon(2n-1)-3y^2+3y-\frac{6x^2y}{2x-1}+\frac{8(n-1)^2 \epsilon xy^2}{2x-1}\notag\\-\frac{3xy}{2x-1}
\end{align}
Here($'$) denotes the derivative with respect to $\eta=\ln a$.
To study this model,we consider the value of deceleration parameter $q_{0}=-0.55$ \cite{na,ab,vm}
From the above autonomous system of differential equations, we find the critical points given in table-V.
	
			\begin{table}[h!]
		
		\centering
			
		\begin{tabular}{|c|c|c|}
			\hline
		\textbf{No}& \textbf{Point} & \textbf{$(x,y)$}\\ \hline
		1 & A & $(0,0)$\\ \hline
		2 & B & $(0,\frac{1}{3}(3.45-0.9n))$\\ \hline
		3 & C & $(0.45,0)$ \\ \hline
		
			\end{tabular}
			
			\caption{Critical point corresponding to autonomous system (28),(29).}
			\end{table}
			At those aforesaid critical points, we construct the linearized Jacobian matrix and corresponding to the Jacobian matrix, we obtain the eigen values and respective values of equation of state parameter $\omega$, given in table-VI.
		\begin{table}[h!]
		
		\resizebox{\linewidth}{!}{	
	
		\begin{tabular}{|c|c|c|}
			\hline
		\textbf{Point}& \textbf{Eigen Value} & \textbf{$\omega$}\\ \hline
		 A & $(1.35,3.45-0.9n)$ & -1\\ \hline
		B & $(-3.45+0.9n,-6.24+10.26n-6.3n^2+1.08n^3)$ & $-1+0.4(3.45-0.9n)(n-1)^2$\\ \hline
		 C & $(-1.35,2.1-0.9n)$ & -1 \\ \hline
		
			\end{tabular}
			}
			\caption{Eigen value corresponding to critical points A,B and C and respective value of $\omega$ at those points.}
			\end{table}	
			
			Here from Table-VI we get that $A$ is a unstable node while $B$ becomes stable node when $n< 3.68$ and $C$ represents a stable node if $n>2.33$.\\
			Since corresponding to the critical point $B$,$\omega$  depends on the value of $n$ and it becomes stable for $n< 3.68$,if we choose $n=3.6$, $\omega=-0.43$. 
\begin{figure}[h!]
\includegraphics[scale=0.5]{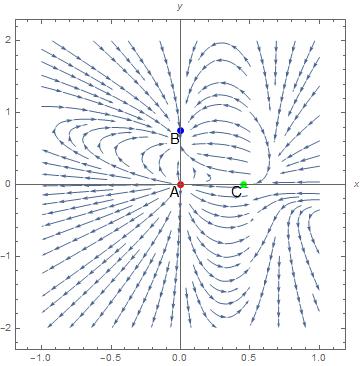}
\caption{Phase plot presenting the behavior of the trajectories for the model in IV B for  $n=2$ and $q_{0}=-0.55$ which shows that A is unstable,B is stable and C is a saddle node for the aforesaid values of the parameter.}
\end{figure}
	\begin{figure}[h!]
\includegraphics[scale=0.5]{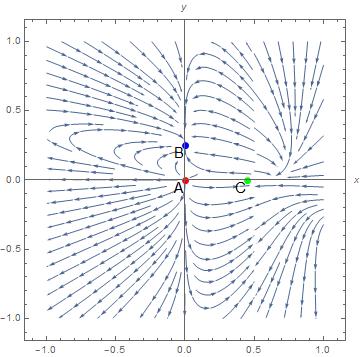}
\caption{Phase plot presenting the behavior of the trajectories for the model in Subsection-B for  $n=3$ and $q_{0}=-0.55$ which shows that A is unstable,B is stable and C is stable node for the aforesaid values of the parameter.}
\end{figure}		  
\section{\label{sec:level V}Conclusion:\protect}
	In this paper, we have considered a cosmological model incorporating modified gravity through non-metricity $Q$ as gravitational interaction.We have employed symmetric teleparallel $f(Q)$ gravity and instead of considering action for normal $f(Q)$ gravity, we have taken the functional form of $f(Q)$ as a combination of two functions of non-metricity $Q$, namely $f_1(Q)=\alpha Q^n$ and $f_2(Q)=Q$ where $f_2(Q)=Q$ is coupled with the Lagrangian matter. We have used dynamical system approach for studying this 
$f(Q)$ gravity model to find out viable solution with late time acceleration.\\
In section IV A, we have considered two values of deceleration parameter $q$ which is close to $-0.5$ for three set of critical points with varying $n$. We observe that, stability of critical points is not affected by the values of $\alpha$. So, we ignore $\alpha$ in our analysis. Though we have used two different values of $q$, but we see that in this chosen range of value, our results vary insignificantly.\\
From table-II, we see that critical point $(0,0)$ is a stable node[Phase plot in Fig-1] for any value of $n\geq4$($P_3,P_4,P_5,P_8,P_9,P_{10}$ in table-II) and corresponding equation of state parameter is $-0.63<\omega$, so $\omega$ lies in the admissible range $-1<\omega<-\frac{1}{3}$ representing quintessence behavior.\\
Critical point $(\frac{1.22\epsilon+\omega}{\omega},0)$ is stable node and represents observed accelerated expansion and the value of $\omega$ implies phantom behavior for $n=2$($P_{11},P_{16}$ in table-III). This node represents an accelerated universe for all other values of $n$($P_{12},P_{13},P_{14},P_{15},P_{17},P_{18},P_{19},P_{20}$ in table-III) and it indicates quintessence behavior for these $n$. So, this critical point is very interesting and can represent a viable solution with late time acceleration.\\
Critical point  $(0,0.33\epsilon-0.67\epsilon n +1)$ is stable node only for  $n=2$ (Phase plot in Fig-6)and $n=3$($P_{21},P_{22},P_{25},P_{26}$ in table-IV) and it may represent accelerated expansion depending on $\omega$.\\
In section IV B, we express the equation of state parameter in terms of dimensionless variables $x,y$. But here $\omega$ is also dependent on $n$ and $\epsilon$, so on $q$. Here also, we get three critical points for particular $q$ and $n$ given in table-V.\\
Though $A$ is unstable, but $\omega$ being -1, represents late time acceleration. Again critical point $B$, $C$ both are stable with late time acceleration. Point $C$ may represent $\Lambda$CDM model, $\omega=-1$.\\ \\
Here, for this paper we have not considered any particular fluid like radiation, baryon, dark energy or dust, rather we focused on general description of fluid. Depending on $\omega$, the dynamics of universe changes and we can say that this work incorporates late time acceleration very well. So, we can see that $f(Q)$ gravity models can be thought as alternative to $\Lambda$CDM model.
\bibliography{Paper}

\providecommand{\noopsort}[1]{}\providecommand{\singleletter}[1]{#1}%
\begin{thebibliography}{59}%
\makeatletter
\providecommand \@ifxundefined [1]{%
 \@ifx{#1\undefined}
}%
\providecommand \@ifnum [1]{%
 \ifnum #1\expandafter \@firstoftwo
 \else \expandafter \@secondoftwo
 \fi
}%
\providecommand \@ifx [1]{%
 \ifx #1\expandafter \@firstoftwo
 \else \expandafter \@secondoftwo
 \fi
}%
\providecommand \natexlab [1]{#1}%
\providecommand \enquote  [1]{``#1''}%
\providecommand \bibnamefont  [1]{#1}%
\providecommand \bibfnamefont [1]{#1}%
\providecommand \citenamefont [1]{#1}%
\providecommand \href@noop [0]{\@secondoftwo}%
\providecommand \href [0]{\begingroup \@sanitize@url \@href}%
\providecommand \@href[1]{\@@startlink{#1}\@@href}%
\providecommand \@@href[1]{\endgroup#1\@@endlink}%
\providecommand \@sanitize@url [0]{\catcode `\\12\catcode `\$12\catcode
  `\&12\catcode `\#12\catcode `\^12\catcode `\_12\catcode `\%12\relax}%
\providecommand \@@startlink[1]{}%
\providecommand \@@endlink[0]{}%
\providecommand \url  [0]{\begingroup\@sanitize@url \@url }%
\providecommand \@url [1]{\endgroup\@href {#1}{\urlprefix }}%
\providecommand \urlprefix  [0]{URL }%
\providecommand \Eprint [0]{\href }%
\providecommand \doibase [0]{https://doi.org/}%
\providecommand \selectlanguage [0]{\@gobble}%
\providecommand \bibinfo  [0]{\@secondoftwo}%
\providecommand \bibfield  [0]{\@secondoftwo}%
\providecommand \translation [1]{[#1]}%
\providecommand \BibitemOpen [0]{}%
\providecommand \bibitemStop [0]{}%
\providecommand \bibitemNoStop [0]{.\EOS\space}%
\providecommand \EOS [0]{\spacefactor3000\relax}%
\providecommand \BibitemShut  [1]{\csname bibitem#1\endcsname}%
\let\auto@bib@innerbib\@empty
%</preamble>
\bibitem [{\citenamefont {A.G.Riess}(2004)}]{ag}%
  \BibitemOpen
  \bibfield  {author} {\bibinfo {author} {\bibnamefont {A.G.Riess}},\
  }\href@noop {} {\bibfield  {journal} {\bibinfo  {journal} {Astrophysics. J.}\
  }\textbf {\bibinfo {volume} {607}},\ \bibinfo {pages} {665} (\bibinfo {year}
  {2004})}\BibitemShut {NoStop}%
\bibitem [{\citenamefont {Perlmutter}\ \emph {et~al.}(1999)\citenamefont
  {Perlmutter}, \citenamefont {Aldering}, \citenamefont {Goldhaber},
  \citenamefont {Knop}, \citenamefont {Nugent} \emph {et~al.}}]{pe}%
  \BibitemOpen
  \bibfield  {author} {\bibinfo {author} {\bibfnamefont {S.}~\bibnamefont
  {Perlmutter}}, \bibinfo {author} {\bibfnamefont {G.}~\bibnamefont
  {Aldering}}, \bibinfo {author} {\bibfnamefont {G.}~\bibnamefont {Goldhaber}},
  \bibinfo {author} {\bibfnamefont {R.~A.}\ \bibnamefont {Knop}}, \bibinfo
  {author} {\bibfnamefont {P.}~\bibnamefont {Nugent}}, \emph {et~al.},\
  }\href@noop {} {\bibfield  {journal} {\bibinfo  {journal} {Astrophysics. J.}\
  }\textbf {\bibinfo {volume} {517}},\ \bibinfo {pages} {565} (\bibinfo {year}
  {1999})}\BibitemShut {NoStop}%
\bibitem [{\citenamefont {V.Sahni}\ and\ \citenamefont
  {A.Starobinsky}(2000)}]{sah}%
  \BibitemOpen
  \bibfield  {author} {\bibinfo {author} {\bibnamefont {V.Sahni}}\ and\
  \bibinfo {author} {\bibnamefont {A.Starobinsky}},\ }\href@noop {} {\bibfield
  {journal} {\bibinfo  {journal} {Int.J.Mod.Phys.D}\ }\textbf {\bibinfo
  {volume} {D9}},\ \bibinfo {pages} {373} (\bibinfo {year} {2000})}\BibitemShut
  {NoStop}%
\bibitem [{\citenamefont {Lobo}(2009)}]{fs}%
  \BibitemOpen
  \bibfield  {author} {\bibinfo {author} {\bibfnamefont {F.~S.~N.}\
  \bibnamefont {Lobo}},\ }\href@noop {} {\bibfield  {journal} {\bibinfo
  {journal} {Research Signpost, ISBN 978-81- 308-0341-8}\ }\textbf {\bibinfo
  {volume} {173}} (\bibinfo {year} {2009})}\BibitemShut {NoStop}%
\bibitem [{\citenamefont {Capozziello}\ and\ \citenamefont
  {Laurentis}(2011)}]{sc}%
  \BibitemOpen
  \bibfield  {author} {\bibinfo {author} {\bibfnamefont {S.}~\bibnamefont
  {Capozziello}}\ and\ \bibinfo {author} {\bibfnamefont {M.~D.}\ \bibnamefont
  {Laurentis}},\ }\href@noop {} {\bibfield  {journal} {\bibinfo  {journal}
  {Physics Reports}\ }\textbf {\bibinfo {volume} {509}},\ \bibinfo {pages}
  {167} (\bibinfo {year} {2011})}\BibitemShut {NoStop}%
\bibitem [{\citenamefont {Hayashi}\ and\ \citenamefont
  {Shirafuji}(1979)}]{khay}%
  \BibitemOpen
  \bibfield  {author} {\bibinfo {author} {\bibfnamefont {K.}~\bibnamefont
  {Hayashi}}\ and\ \bibinfo {author} {\bibfnamefont {T.}~\bibnamefont
  {Shirafuji}},\ }\href@noop {} {\bibfield  {journal} {\bibinfo  {journal}
  {Phys. Rev. D.}\ }\textbf {\bibinfo {volume} {19}},\ \bibinfo {pages} {3524}
  (\bibinfo {year} {1979})}\BibitemShut {NoStop}%
\bibitem [{\citenamefont {Maluf}(1994)}]{jwma}%
  \BibitemOpen
  \bibfield  {author} {\bibinfo {author} {\bibfnamefont {J.}~\bibnamefont
  {Maluf}},\ }\href@noop {} {\bibfield  {journal} {\bibinfo  {journal} {J.
  Math. Phys.}\ }\textbf {\bibinfo {volume} {35}},\ \bibinfo {pages} {335}
  (\bibinfo {year} {1994})}\BibitemShut {NoStop}%
\bibitem [{\citenamefont {Arcos}\ and\ \citenamefont {Pereira}(2004)}]{hiar}%
  \BibitemOpen
  \bibfield  {author} {\bibinfo {author} {\bibfnamefont {H.}~\bibnamefont
  {Arcos}}\ and\ \bibinfo {author} {\bibfnamefont {J.}~\bibnamefont
  {Pereira}},\ }\href@noop {} {\bibfield  {journal} {\bibinfo  {journal} {Int.
  J. Mod. Phys. D}\ }\textbf {\bibinfo {volume} {13}},\ \bibinfo {pages} {2193}
  (\bibinfo {year} {2004})}\BibitemShut {NoStop}%
\bibitem [{\citenamefont {Astier}(2006)}]{as}%
  \BibitemOpen
  \bibfield  {author} {\bibinfo {author} {\bibfnamefont {P.}~\bibnamefont
  {Astier}},\ }\href@noop {} {\bibfield  {journal} {\bibinfo  {journal}
  {Astrophysics. J.}\ }\textbf {\bibinfo {volume} {31}},\ \bibinfo {pages}
  {447} (\bibinfo {year} {2006})}\BibitemShut {NoStop}%
\bibitem [{\citenamefont {Spergel}(2007)}]{sp}%
  \BibitemOpen
  \bibfield  {author} {\bibinfo {author} {\bibfnamefont {D.~N.}\ \bibnamefont
  {Spergel}},\ }\href@noop {} {\bibfield  {journal} {\bibinfo  {journal}
  {Astrophysics. J. Suppl.}\ }\textbf {\bibinfo {volume} {170}},\ \bibinfo
  {pages} {377} (\bibinfo {year} {2007})}\BibitemShut {NoStop}%
\bibitem [{\citenamefont {Eisenstein}\ \emph {et~al.}(2005)\citenamefont
  {Eisenstein}, \citenamefont {zehavi}, \citenamefont {Hogg}, \citenamefont
  {Scoccimarro}, \citenamefont {Blanton} \emph {et~al.}}]{dj}%
  \BibitemOpen
  \bibfield  {author} {\bibinfo {author} {\bibfnamefont {D.~J.}\ \bibnamefont
  {Eisenstein}}, \bibinfo {author} {\bibfnamefont {I.}~\bibnamefont {zehavi}},
  \bibinfo {author} {\bibfnamefont {D.~W.}\ \bibnamefont {Hogg}}, \bibinfo
  {author} {\bibfnamefont {R.}~\bibnamefont {Scoccimarro}}, \bibinfo {author}
  {\bibfnamefont {M.~R.}\ \bibnamefont {Blanton}}, \emph {et~al.},\ }\href@noop
  {} {\bibfield  {journal} {\bibinfo  {journal} {Astrophys. J.}\ }\textbf
  {\bibinfo {volume} {633}},\ \bibinfo {pages} {560} (\bibinfo {year}
  {2005})}\BibitemShut {NoStop}%
\bibitem [{\citenamefont {Wald}(1984)}]{ld}%
  \BibitemOpen
  \bibfield  {author} {\bibinfo {author} {\bibfnamefont {R.~M.}\ \bibnamefont
  {Wald}},\ }\bibinfo {title} {General relativity}\ (\bibinfo  {publisher}
  {University of Chicago Press},\ \bibinfo {address} {USA},\ \bibinfo {year}
  {1984})\BibitemShut {NoStop}%
\bibitem [{\citenamefont {Sahni}\ \emph {et~al.}(2003)\citenamefont {Sahni},
  \citenamefont {Saini}, \citenamefont {Starobinsky},\ and\ \citenamefont
  {Alam}}]{td}%
  \BibitemOpen
  \bibfield  {author} {\bibinfo {author} {\bibfnamefont {V.}~\bibnamefont
  {Sahni}}, \bibinfo {author} {\bibfnamefont {T.~D.}\ \bibnamefont {Saini}},
  \bibinfo {author} {\bibfnamefont {A.~A.}\ \bibnamefont {Starobinsky}},\ and\
  \bibinfo {author} {\bibfnamefont {U.}~\bibnamefont {Alam}},\ }\href@noop {}
  {\bibfield  {journal} {\bibinfo  {journal} {Journal of Experimental and
  Theoretical Physics Letters}\ }\textbf {\bibinfo {volume} {77}},\ \bibinfo
  {pages} {201} (\bibinfo {year} {2003})}\BibitemShut {NoStop}%
\bibitem [{\citenamefont {Sahni}(2004)}]{ni}%
  \BibitemOpen
  \bibfield  {author} {\bibinfo {author} {\bibfnamefont {V.}~\bibnamefont
  {Sahni}},\ }\href@noop {} {\bibfield  {journal} {\bibinfo  {journal} {The
  Physics of the Early Universe}\ ,\ \bibinfo {pages} {141}} (\bibinfo {year}
  {2004})}\BibitemShut {NoStop}%
\bibitem [{\citenamefont {Jungman}\ \emph {et~al.}(1996)\citenamefont
  {Jungman}, \citenamefont {Kamionkowski},\ and\ \citenamefont {Griest}}]{gj}%
  \BibitemOpen
  \bibfield  {author} {\bibinfo {author} {\bibfnamefont {G.}~\bibnamefont
  {Jungman}}, \bibinfo {author} {\bibfnamefont {M.}~\bibnamefont
  {Kamionkowski}},\ and\ \bibinfo {author} {\bibfnamefont {K.}~\bibnamefont
  {Griest}},\ }\href@noop {} {\bibfield  {journal} {\bibinfo  {journal}
  {Physics Reports}\ }\textbf {\bibinfo {volume} {267}},\ \bibinfo {pages}
  {195} (\bibinfo {year} {1996})}\BibitemShut {NoStop}%
\bibitem [{\citenamefont {Zlatev}\ \emph {et~al.}(1999)\citenamefont {Zlatev},
  \citenamefont {Wang},\ and\ \citenamefont {Steinhardt}}]{iz}%
  \BibitemOpen
  \bibfield  {author} {\bibinfo {author} {\bibfnamefont {I.}~\bibnamefont
  {Zlatev}}, \bibinfo {author} {\bibfnamefont {L.}~\bibnamefont {Wang}},\ and\
  \bibinfo {author} {\bibfnamefont {P.~J.}\ \bibnamefont {Steinhardt}},\
  }\href@noop {} {\bibfield  {journal} {\bibinfo  {journal} {Phys. Rev. Lett.}\
  }\textbf {\bibinfo {volume} {82}},\ \bibinfo {pages} {896} (\bibinfo {year}
  {1999})}\BibitemShut {NoStop}%
\bibitem [{\citenamefont {Chimento}\ and\ \citenamefont
  {Feinstein}(2020)}]{cf}%
  \BibitemOpen
  \bibfield  {author} {\bibinfo {author} {\bibfnamefont {L.~P.}\ \bibnamefont
  {Chimento}}\ and\ \bibinfo {author} {\bibfnamefont {A.}~\bibnamefont
  {Feinstein}},\ }\href@noop {} {\bibfield  {journal} {\bibinfo  {journal}
  {Modern Physics Letters A}\ }\textbf {\bibinfo {volume} {19}},\ \bibinfo
  {pages} {761} (\bibinfo {year} {2020})}\BibitemShut {NoStop}%
\bibitem [{\citenamefont {Nojiri}\ and\ \citenamefont {Odintsov}(2007)}]{ss}%
  \BibitemOpen
  \bibfield  {author} {\bibinfo {author} {\bibfnamefont {S.}~\bibnamefont
  {Nojiri}}\ and\ \bibinfo {author} {\bibfnamefont {S.~D.}\ \bibnamefont
  {Odintsov}},\ }\href@noop {} {\bibfield  {journal} {\bibinfo  {journal}
  {International Journal of Geometric Methods in Modern Physics}\ }\textbf
  {\bibinfo {volume} {04}},\ \bibinfo {pages} {115} (\bibinfo {year}
  {2007})}\BibitemShut {NoStop}%
\bibitem [{\citenamefont {F.K.Anagnostopoulos}\ \emph
  {et~al.}(2021)\citenamefont {F.K.Anagnostopoulos}, \citenamefont
  {Basilakos},\ and\ \citenamefont {E.N.Saridakis}}]{cha}%
  \BibitemOpen
  \bibfield  {author} {\bibinfo {author} {\bibnamefont {F.K.Anagnostopoulos}},
  \bibinfo {author} {\bibfnamefont {S.}~\bibnamefont {Basilakos}},\ and\
  \bibinfo {author} {\bibnamefont {E.N.Saridakis}},\ }\href@noop {} {\bibfield
  {journal} {\bibinfo  {journal} {Phys Letters B}\ }\textbf {\bibinfo {volume}
  {822}},\ \bibinfo {pages} {136634} (\bibinfo {year} {2021})}\BibitemShut
  {NoStop}%
\bibitem [{\citenamefont {Nojiri}\ \emph {et~al.}(2017)\citenamefont {Nojiri},
  \citenamefont {Odintsov},\ and\ \citenamefont {Oikonomou}}]{sn}%
  \BibitemOpen
  \bibfield  {author} {\bibinfo {author} {\bibfnamefont {S.}~\bibnamefont
  {Nojiri}}, \bibinfo {author} {\bibfnamefont {S.~D.}\ \bibnamefont
  {Odintsov}},\ and\ \bibinfo {author} {\bibfnamefont {V.~K.}\ \bibnamefont
  {Oikonomou}},\ }\href@noop {} {\bibfield  {journal} {\bibinfo  {journal}
  {Phys. Rept.}\ }\textbf {\bibinfo {volume} {692}},\ \bibinfo {pages} {1}
  (\bibinfo {year} {2017})}\BibitemShut {NoStop}%
\bibitem [{\citenamefont {B{\"o}hmer}\ and\ \citenamefont {Jensko}(2021)}]{bo}%
  \BibitemOpen
  \bibfield  {author} {\bibinfo {author} {\bibfnamefont {C.~G.}\ \bibnamefont
  {B{\"o}hmer}}\ and\ \bibinfo {author} {\bibfnamefont {E.}~\bibnamefont
  {Jensko}},\ }\href@noop {} {\bibfield  {journal} {\bibinfo  {journal} {Phys.
  Rev. D.}\ }\textbf {\bibinfo {volume} {104}},\ \bibinfo {pages} {024010}
  (\bibinfo {year} {2021})}\BibitemShut {NoStop}%
\bibitem [{\citenamefont {Braglia}\ \emph {et~al.}(2021)\citenamefont
  {Braglia}, \citenamefont {Ballardini}, \citenamefont {Finelli},\ and\
  \citenamefont {Koyama}}]{bra}%
  \BibitemOpen
  \bibfield  {author} {\bibinfo {author} {\bibfnamefont {M.}~\bibnamefont
  {Braglia}}, \bibinfo {author} {\bibfnamefont {M.}~\bibnamefont {Ballardini}},
  \bibinfo {author} {\bibfnamefont {F.}~\bibnamefont {Finelli}},\ and\ \bibinfo
  {author} {\bibfnamefont {K.}~\bibnamefont {Koyama}},\ }\href@noop {}
  {\bibfield  {journal} {\bibinfo  {journal} {Phys. Rev. D.}\ }\textbf
  {\bibinfo {volume} {103}},\ \bibinfo {pages} {043528} (\bibinfo {year}
  {2021})}\BibitemShut {NoStop}%
\bibitem [{\citenamefont {Saridakis}\ \emph {et~al.}(2021)\citenamefont
  {Saridakis}, \citenamefont {Lazkoz}, \citenamefont {Salzano}, \citenamefont
  {Moniz}, \citenamefont {Capozziello} \emph {et~al.}}]{ens}%
  \BibitemOpen
  \bibfield  {author} {\bibinfo {author} {\bibfnamefont {E.~N.}\ \bibnamefont
  {Saridakis}}, \bibinfo {author} {\bibfnamefont {R.}~\bibnamefont {Lazkoz}},
  \bibinfo {author} {\bibfnamefont {V.}~\bibnamefont {Salzano}}, \bibinfo
  {author} {\bibfnamefont {P.~V.}\ \bibnamefont {Moniz}}, \bibinfo {author}
  {\bibfnamefont {S.}~\bibnamefont {Capozziello}}, \emph {et~al.},\ }\bibinfo
  {title} {Modified gravity and cosmology: An update by the cantata network}\
  (\bibinfo  {publisher} {Springer},\ \bibinfo {address} {Berlin},\ \bibinfo
  {year} {2021})\BibitemShut {NoStop}%
\bibitem [{\citenamefont {Mussatayeva}\ \emph {et~al.}(2023)\citenamefont
  {Mussatayeva}, \citenamefont {Myrzakulov},\ and\ \citenamefont
  {Koussour}}]{amus}%
  \BibitemOpen
  \bibfield  {author} {\bibinfo {author} {\bibfnamefont {A.}~\bibnamefont
  {Mussatayeva}}, \bibinfo {author} {\bibfnamefont {N.}~\bibnamefont
  {Myrzakulov}},\ and\ \bibinfo {author} {\bibfnamefont {M.}~\bibnamefont
  {Koussour}},\ }\href@noop {} {\bibfield  {journal} {\bibinfo  {journal}
  {Physics of the Dark Universe}\ }\textbf {\bibinfo {volume} {47}},\ \bibinfo
  {pages} {101276} (\bibinfo {year} {2023})}\BibitemShut {NoStop}%
\bibitem [{\citenamefont {Myrzakulov}\ \emph {et~al.}(2023)\citenamefont
  {Myrzakulov}, \citenamefont {Koussour},\ and\ \citenamefont {Gogoi}}]{nm}%
  \BibitemOpen
  \bibfield  {author} {\bibinfo {author} {\bibfnamefont {N.}~\bibnamefont
  {Myrzakulov}}, \bibinfo {author} {\bibfnamefont {M.}~\bibnamefont
  {Koussour}},\ and\ \bibinfo {author} {\bibfnamefont {D.~J.}\ \bibnamefont
  {Gogoi}},\ }\href@noop {} {\bibfield  {journal} {\bibinfo  {journal} {Physics
  of the Dark Universe}\ }\textbf {\bibinfo {volume} {47}},\ \bibinfo {pages}
  {101268} (\bibinfo {year} {2023})}\BibitemShut {NoStop}%
\bibitem [{\citenamefont {Jimenez}\ \emph {et~al.}(2018)\citenamefont
  {Jimenez}, \citenamefont {L.Heisenberg},\ and\ \citenamefont
  {T.Koivisto}}]{nonmet}%
  \BibitemOpen
  \bibfield  {author} {\bibinfo {author} {\bibfnamefont {J.}~\bibnamefont
  {Jimenez}}, \bibinfo {author} {\bibnamefont {L.Heisenberg}},\ and\ \bibinfo
  {author} {\bibnamefont {T.Koivisto}},\ }\href@noop {} {\bibfield  {journal}
  {\bibinfo  {journal} {Phys. Rev.D}\ }\textbf {\bibinfo {volume} {98}},\
  \bibinfo {pages} {044048} (\bibinfo {year} {2018})}\BibitemShut {NoStop}%
\bibitem [{\citenamefont {Albuquerque}\ and\ \citenamefont
  {Frusciante}(2022)}]{is}%
  \BibitemOpen
  \bibfield  {author} {\bibinfo {author} {\bibfnamefont {I.~S.}\ \bibnamefont
  {Albuquerque}}\ and\ \bibinfo {author} {\bibfnamefont {N.}~\bibnamefont
  {Frusciante}},\ }\href@noop {} {\bibfield  {journal} {\bibinfo  {journal}
  {Physics of the Dark Universe}\ }\textbf {\bibinfo {volume} {35}},\ \bibinfo
  {pages} {100980} (\bibinfo {year} {2022})}\BibitemShut {NoStop}%
\bibitem [{\citenamefont {Lazkoz}\ \emph {et~al.}(2019)\citenamefont {Lazkoz},
  \citenamefont {Lobo}, \citenamefont {Banos},\ and\ \citenamefont
  {Salzano}}]{la}%
  \BibitemOpen
  \bibfield  {author} {\bibinfo {author} {\bibfnamefont {R.}~\bibnamefont
  {Lazkoz}}, \bibinfo {author} {\bibfnamefont {F.~S.~N.}\ \bibnamefont {Lobo}},
  \bibinfo {author} {\bibfnamefont {M.~O.}\ \bibnamefont {Banos}},\ and\
  \bibinfo {author} {\bibfnamefont {V.}~\bibnamefont {Salzano}},\ }\href@noop
  {} {\bibfield  {journal} {\bibinfo  {journal} {Phys. Rev. D}\ }\textbf
  {\bibinfo {volume} {100}},\ \bibinfo {pages} {104027} (\bibinfo {year}
  {2019})}\BibitemShut {NoStop}%
\bibitem [{\citenamefont {Atayde}\ and\ \citenamefont {Frusciante}(2021)}]{ta}%
  \BibitemOpen
  \bibfield  {author} {\bibinfo {author} {\bibfnamefont {L.}~\bibnamefont
  {Atayde}}\ and\ \bibinfo {author} {\bibfnamefont {N.}~\bibnamefont
  {Frusciante}},\ }\href@noop {} {\bibfield  {journal} {\bibinfo  {journal}
  {Phys. Rev. D.}\ }\textbf {\bibinfo {volume} {104}},\ \bibinfo {pages}
  {064052} (\bibinfo {year} {2021})}\BibitemShut {NoStop}%
\bibitem [{\citenamefont {Ayuso}\ \emph {et~al.}(2021)\citenamefont {Ayuso},
  \citenamefont {Lazkoz},\ and\ \citenamefont {Salzano}}]{yu}%
  \BibitemOpen
  \bibfield  {author} {\bibinfo {author} {\bibfnamefont {I.}~\bibnamefont
  {Ayuso}}, \bibinfo {author} {\bibfnamefont {R.}~\bibnamefont {Lazkoz}},\ and\
  \bibinfo {author} {\bibfnamefont {V.}~\bibnamefont {Salzano}},\ }\href@noop
  {} {\bibfield  {journal} {\bibinfo  {journal} {Phys. Rev. D.}\ }\textbf
  {\bibinfo {volume} {103}},\ \bibinfo {pages} {063505} (\bibinfo {year}
  {2021})}\BibitemShut {NoStop}%
\bibitem [{\citenamefont {Frusciante}(2021)}]{nf}%
  \BibitemOpen
  \bibfield  {author} {\bibinfo {author} {\bibfnamefont {N.}~\bibnamefont
  {Frusciante}},\ }\href@noop {} {\bibfield  {journal} {\bibinfo  {journal}
  {Phys. Rev. D.}\ }\textbf {\bibinfo {volume} {103}},\ \bibinfo {pages}
  {044021} (\bibinfo {year} {2021})}\BibitemShut {NoStop}%
\bibitem [{\citenamefont {Mandal}\ and\ \citenamefont {Sahoo}(2021)}]{sm}%
  \BibitemOpen
  \bibfield  {author} {\bibinfo {author} {\bibfnamefont {S.}~\bibnamefont
  {Mandal}}\ and\ \bibinfo {author} {\bibfnamefont {P.}~\bibnamefont {Sahoo}},\
  }\href@noop {} {\bibfield  {journal} {\bibinfo  {journal} {Phys. Letters. B}\
  }\textbf {\bibinfo {volume} {823}},\ \bibinfo {pages} {136786} (\bibinfo
  {year} {2021})}\BibitemShut {NoStop}%
\bibitem [{\citenamefont {Shabani}\ \emph {et~al.}(2023)\citenamefont
  {Shabani}, \citenamefont {De},\ and\ \citenamefont {Loo}}]{ab}%
  \BibitemOpen
  \bibfield  {author} {\bibinfo {author} {\bibfnamefont {H.}~\bibnamefont
  {Shabani}}, \bibinfo {author} {\bibfnamefont {A.}~\bibnamefont {De}},\ and\
  \bibinfo {author} {\bibfnamefont {T.}~\bibnamefont {Loo}},\ }\href@noop {}
  {\bibfield  {journal} {\bibinfo  {journal} {Eur. Phys. J. C}\ }\textbf
  {\bibinfo {volume} {83}},\ \bibinfo {pages} {535} (\bibinfo {year}
  {2023})}\BibitemShut {NoStop}%
\bibitem [{\citenamefont {Narawade}\ \emph {et~al.}(2032)\citenamefont
  {Narawade}, \citenamefont {Singh},\ and\ \citenamefont {Mishra}}]{sana}%
  \BibitemOpen
  \bibfield  {author} {\bibinfo {author} {\bibfnamefont {S.~A.}\ \bibnamefont
  {Narawade}}, \bibinfo {author} {\bibfnamefont {S.~P.}\ \bibnamefont
  {Singh}},\ and\ \bibinfo {author} {\bibfnamefont {B.}~\bibnamefont
  {Mishra}},\ }\href@noop {} {\bibfield  {journal} {\bibinfo  {journal}
  {Physics of the Dark Universe}\ }\textbf {\bibinfo {volume} {42}},\ \bibinfo
  {pages} {101282} (\bibinfo {year} {2032})}\BibitemShut {NoStop}%
\bibitem [{\citenamefont {Khyllep}\ \emph {et~al.}(2021)\citenamefont
  {Khyllep}, \citenamefont {Paliathanasis},\ and\ \citenamefont {Dutta}}]{wk}%
  \BibitemOpen
  \bibfield  {author} {\bibinfo {author} {\bibfnamefont {W.}~\bibnamefont
  {Khyllep}}, \bibinfo {author} {\bibfnamefont {A.}~\bibnamefont
  {Paliathanasis}},\ and\ \bibinfo {author} {\bibfnamefont {J.}~\bibnamefont
  {Dutta}},\ }\href@noop {} {\bibfield  {journal} {\bibinfo  {journal} {Phys.
  Rev. D.}\ }\textbf {\bibinfo {volume} {103}},\ \bibinfo {pages} {103521}
  (\bibinfo {year} {2021})}\BibitemShut {NoStop}%
\bibitem [{\citenamefont {Nester}\ and\ \citenamefont {Yo}(1999)}]{jm}%
  \BibitemOpen
  \bibfield  {author} {\bibinfo {author} {\bibfnamefont {J.~M.}\ \bibnamefont
  {Nester}}\ and\ \bibinfo {author} {\bibfnamefont {H.~J.}\ \bibnamefont
  {Yo}},\ }\href@noop {} {\bibfield  {journal} {\bibinfo  {journal}
  {Chin.J.Phys.}\ }\textbf {\bibinfo {volume} {37}},\ \bibinfo {pages} {113}
  (\bibinfo {year} {1999})}\BibitemShut {NoStop}%
\bibitem [{\citenamefont {Adak}\ and\ \citenamefont {Sert}(2005)}]{ad}%
  \BibitemOpen
  \bibfield  {author} {\bibinfo {author} {\bibfnamefont {M.}~\bibnamefont
  {Adak}}\ and\ \bibinfo {author} {\bibfnamefont {{\"O}.}~\bibnamefont
  {Sert}},\ }\href@noop {} {\bibfield  {journal} {\bibinfo  {journal} {Turk. J.
  Phys.}\ }\textbf {\bibinfo {volume} {29}} (\bibinfo {year}
  {2005})}\BibitemShut {NoStop}%
\bibitem [{\citenamefont {Maluf}(2013)}]{jw}%
  \BibitemOpen
  \bibfield  {author} {\bibinfo {author} {\bibfnamefont {J.~W.}\ \bibnamefont
  {Maluf}},\ }\href@noop {} {\bibfield  {journal} {\bibinfo  {journal} {Annalen
  Phys.}\ }\textbf {\bibinfo {volume} {525}},\ \bibinfo {pages} {339} (\bibinfo
  {year} {2013})}\BibitemShut {NoStop}%
\bibitem [{\citenamefont {Flathmann}\ and\ \citenamefont {Hohmann}(2021)}]{kf}%
  \BibitemOpen
  \bibfield  {author} {\bibinfo {author} {\bibfnamefont {K.}~\bibnamefont
  {Flathmann}}\ and\ \bibinfo {author} {\bibfnamefont {M.}~\bibnamefont
  {Hohmann}},\ }\href@noop {} {\bibfield  {journal} {\bibinfo  {journal} {Phys.
  Rev. D.}\ }\textbf {\bibinfo {volume} {103}},\ \bibinfo {pages} {044030}
  (\bibinfo {year} {2021})}\BibitemShut {NoStop}%
\bibitem [{\citenamefont {Ferraro}\ and\ \citenamefont {Fiorini}(2007)}]{rfe}%
  \BibitemOpen
  \bibfield  {author} {\bibinfo {author} {\bibfnamefont {R.}~\bibnamefont
  {Ferraro}}\ and\ \bibinfo {author} {\bibfnamefont {F.}~\bibnamefont
  {Fiorini}},\ }\href@noop {} {\bibfield  {journal} {\bibinfo  {journal} {Phys.
  Rev. D.}\ }\textbf {\bibinfo {volume} {75}},\ \bibinfo {pages} {084031}
  (\bibinfo {year} {2007})}\BibitemShut {NoStop}%
\bibitem [{\citenamefont {Lu}\ \emph {et~al.}(2019)\citenamefont {Lu},
  \citenamefont {Zhao},\ and\ \citenamefont {Chee}}]{jl}%
  \BibitemOpen
  \bibfield  {author} {\bibinfo {author} {\bibfnamefont {J.}~\bibnamefont
  {Lu}}, \bibinfo {author} {\bibfnamefont {X.}~\bibnamefont {Zhao}},\ and\
  \bibinfo {author} {\bibfnamefont {G.}~\bibnamefont {Chee}},\ }\href@noop {}
  {\bibfield  {journal} {\bibinfo  {journal} {Eur. Phys. J. C}\ }\textbf
  {\bibinfo {volume} {79}},\ \bibinfo {pages} {530} (\bibinfo {year}
  {2019})}\BibitemShut {NoStop}%
\bibitem [{\citenamefont {Narawade}\ \emph {et~al.}(2022)\citenamefont
  {Narawade}, \citenamefont {L.Pati}, \citenamefont {Mishra},\ and\
  \citenamefont {Tripathy}}]{san}%
  \BibitemOpen
  \bibfield  {author} {\bibinfo {author} {\bibfnamefont {S.~A.}\ \bibnamefont
  {Narawade}}, \bibinfo {author} {\bibnamefont {L.Pati}}, \bibinfo {author}
  {\bibfnamefont {B.}~\bibnamefont {Mishra}},\ and\ \bibinfo {author}
  {\bibfnamefont {S.~K.}\ \bibnamefont {Tripathy}},\ }\href@noop {} {\bibfield
  {journal} {\bibinfo  {journal} {Physics of the Dark Universe}\ }\textbf
  {\bibinfo {volume} {36}},\ \bibinfo {pages} {101020} (\bibinfo {year}
  {2022})}\BibitemShut {NoStop}%
\bibitem [{\citenamefont {Jimenez}\ \emph {et~al.}(2020)\citenamefont
  {Jimenez}, \citenamefont {L.Heisenberg}, \citenamefont {Koivisto},\ and\
  \citenamefont {Pekar}}]{jhk}%
  \BibitemOpen
  \bibfield  {author} {\bibinfo {author} {\bibfnamefont {J.}~\bibnamefont
  {Jimenez}}, \bibinfo {author} {\bibnamefont {L.Heisenberg}}, \bibinfo
  {author} {\bibfnamefont {T.}~\bibnamefont {Koivisto}},\ and\ \bibinfo
  {author} {\bibfnamefont {S.}~\bibnamefont {Pekar}},\ }\href@noop {}
  {\bibfield  {journal} {\bibinfo  {journal} {Phys. Rev.D}\ }\textbf {\bibinfo
  {volume} {101}},\ \bibinfo {pages} {103507} (\bibinfo {year}
  {2020})}\BibitemShut {NoStop}%
\bibitem [{\citenamefont {J{\"a}rv}\ \emph {et~al.}(2018)\citenamefont
  {J{\"a}rv}, \citenamefont {R{\"u}nkla}, \citenamefont {Saal},\ and\
  \citenamefont {Vilson}}]{lj}%
  \BibitemOpen
  \bibfield  {author} {\bibinfo {author} {\bibfnamefont {L.}~\bibnamefont
  {J{\"a}rv}}, \bibinfo {author} {\bibfnamefont {M.}~\bibnamefont
  {R{\"u}nkla}}, \bibinfo {author} {\bibfnamefont {M.}~\bibnamefont {Saal}},\
  and\ \bibinfo {author} {\bibfnamefont {O.}~\bibnamefont {Vilson}},\
  }\href@noop {} {\bibfield  {journal} {\bibinfo  {journal} {Phys. Rev. D.}\
  }\textbf {\bibinfo {volume} {97}},\ \bibinfo {pages} {124025} (\bibinfo
  {year} {2018})}\BibitemShut {NoStop}%
\bibitem [{\citenamefont {Mol}(2017)}]{im}%
  \BibitemOpen
  \bibfield  {author} {\bibinfo {author} {\bibfnamefont {I.}~\bibnamefont
  {Mol}},\ }\href@noop {} {\bibfield  {journal} {\bibinfo  {journal} {Adv.
  Appl. Clifford Algebras}\ }\textbf {\bibinfo {volume} {3}},\ \bibinfo {pages}
  {2607} (\bibinfo {year} {2017})}\BibitemShut {NoStop}%
\bibitem [{\citenamefont {L.Perko}(2000)}]{lp}%
  \BibitemOpen
  \bibfield  {author} {\bibinfo {author} {\bibnamefont {L.Perko}},\ }\bibinfo
  {title} {Differential equations and dynamical systems}\ (\bibinfo
  {publisher} {Springer},\ \bibinfo {address} {New York},\ \bibinfo {year}
  {2000})\ \bibinfo {edition} {3rd}\ ed.\BibitemShut {Stop}%
\bibitem [{\citenamefont {S.Bahamonde}\ \emph {et~al.}(2018)\citenamefont
  {S.Bahamonde}, \citenamefont {B{\"o}hmer}, \citenamefont {Carloni},
  \citenamefont {Copeland}, \citenamefont {Fang},\ and\ \citenamefont
  {Tamanini}}]{ba}%
  \BibitemOpen
  \bibfield  {author} {\bibinfo {author} {\bibnamefont {S.Bahamonde}}, \bibinfo
  {author} {\bibfnamefont {C.}~\bibnamefont {B{\"o}hmer}}, \bibinfo {author}
  {\bibfnamefont {S.}~\bibnamefont {Carloni}}, \bibinfo {author} {\bibfnamefont
  {E.~J.}\ \bibnamefont {Copeland}}, \bibinfo {author} {\bibfnamefont
  {W.}~\bibnamefont {Fang}},\ and\ \bibinfo {author} {\bibfnamefont
  {N.}~\bibnamefont {Tamanini}},\ }\href@noop {} {\bibfield  {journal}
  {\bibinfo  {journal} {Physics Reports}\ }\textbf {\bibinfo {volume}
  {775-777}} (\bibinfo {year} {2018})}\BibitemShut {NoStop}%
\bibitem [{\citenamefont {Carloni}\ \emph {et~al.}(2016)\citenamefont
  {Carloni}, \citenamefont {Lobo}, \citenamefont {Otalora},\ and\ \citenamefont
  {Saridakis}}]{lo}%
  \BibitemOpen
  \bibfield  {author} {\bibinfo {author} {\bibfnamefont {S.}~\bibnamefont
  {Carloni}}, \bibinfo {author} {\bibfnamefont {F.~S.~N.}\ \bibnamefont
  {Lobo}}, \bibinfo {author} {\bibfnamefont {G.}~\bibnamefont {Otalora}},\ and\
  \bibinfo {author} {\bibfnamefont {E.~N.}\ \bibnamefont {Saridakis}},\
  }\href@noop {} {\bibfield  {journal} {\bibinfo  {journal} {Phys. Rev. D.}\
  }\textbf {\bibinfo {volume} {93}},\ \bibinfo {pages} {024034} (\bibinfo
  {year} {2016})}\BibitemShut {NoStop}%
\bibitem [{\citenamefont {Wainwright}\ and\ \citenamefont
  {Ellis}(1997)}]{jwai}%
  \BibitemOpen
  \bibfield  {author} {\bibinfo {author} {\bibfnamefont {J.}~\bibnamefont
  {Wainwright}}\ and\ \bibinfo {author} {\bibfnamefont {G.~F.~R.}\ \bibnamefont
  {Ellis}},\ }\bibinfo {title} {Dynamical systems in cosmology}\ (\bibinfo
  {publisher} {Cambridge University Press},\ \bibinfo {year}
  {1997})\BibitemShut {NoStop}%
\bibitem [{\citenamefont {Scolnic}\ \emph {et~al.}(2018)\citenamefont
  {Scolnic}, \citenamefont {Jones}, \citenamefont {Rest}, \citenamefont {Pan},
  \citenamefont {Chornock} \emph {et~al.}}]{dm}%
  \BibitemOpen
  \bibfield  {author} {\bibinfo {author} {\bibfnamefont {D.~M.}\ \bibnamefont
  {Scolnic}}, \bibinfo {author} {\bibfnamefont {D.~O.}\ \bibnamefont {Jones}},
  \bibinfo {author} {\bibfnamefont {A.}~\bibnamefont {Rest}}, \bibinfo {author}
  {\bibfnamefont {Y.~C.}\ \bibnamefont {Pan}}, \bibinfo {author} {\bibfnamefont
  {R.}~\bibnamefont {Chornock}}, \emph {et~al.},\ }\href@noop {} {\bibfield
  {journal} {\bibinfo  {journal} {The Astrophysical Journal}\ }\textbf
  {\bibinfo {volume} {859}},\ \bibinfo {pages} {101} (\bibinfo {year}
  {2018})}\BibitemShut {NoStop}%
\bibitem [{\citenamefont {Sharov}\ and\ \citenamefont {Vasiliev}(2018)}]{gs}%
  \BibitemOpen
  \bibfield  {author} {\bibinfo {author} {\bibfnamefont {G.~S.}\ \bibnamefont
  {Sharov}}\ and\ \bibinfo {author} {\bibfnamefont {V.~O.}\ \bibnamefont
  {Vasiliev}},\ }\href@noop {} {\bibfield  {journal} {\bibinfo  {journal}
  {Mathematical Modelling and Geometry}\ }\textbf {\bibinfo {volume} {6}},\
  \bibinfo {pages} {1} (\bibinfo {year} {2018})}\BibitemShut {NoStop}%
\bibitem [{\citenamefont {Aghanim}\ \emph {et~al.}(2020)\citenamefont
  {Aghanim}, \citenamefont {Akrami}, \citenamefont {Ashdown}, \citenamefont
  {Aumont}, \citenamefont {Baccigalupi} \emph {et~al.}}]{na}%
  \BibitemOpen
  \bibfield  {author} {\bibinfo {author} {\bibfnamefont {N.}~\bibnamefont
  {Aghanim}}, \bibinfo {author} {\bibfnamefont {Y.}~\bibnamefont {Akrami}},
  \bibinfo {author} {\bibfnamefont {M.}~\bibnamefont {Ashdown}}, \bibinfo
  {author} {\bibfnamefont {J.}~\bibnamefont {Aumont}}, \bibinfo {author}
  {\bibfnamefont {C.}~\bibnamefont {Baccigalupi}}, \emph {et~al.},\ }\href@noop
  {} {\bibfield  {journal} {\bibinfo  {journal} {Astronomy \& Astrophysics}\
  }\textbf {\bibinfo {volume} {641}},\ \bibinfo {pages} {67} (\bibinfo {year}
  {2020})}\BibitemShut {NoStop}%
\bibitem [{\citenamefont {Delubac}\ \emph {et~al.}(2015)\citenamefont
  {Delubac}, \citenamefont {Bautista}, \citenamefont {Busca}, \citenamefont
  {Rich}, \citenamefont {Kirkby} \emph {et~al.}}]{de}%
  \BibitemOpen
  \bibfield  {author} {\bibinfo {author} {\bibfnamefont {T.}~\bibnamefont
  {Delubac}}, \bibinfo {author} {\bibfnamefont {J.~E.}\ \bibnamefont
  {Bautista}}, \bibinfo {author} {\bibfnamefont {N.~G.}\ \bibnamefont {Busca}},
  \bibinfo {author} {\bibfnamefont {J.}~\bibnamefont {Rich}}, \bibinfo {author}
  {\bibfnamefont {D.}~\bibnamefont {Kirkby}}, \emph {et~al.},\ }\href@noop {}
  {\bibfield  {journal} {\bibinfo  {journal} {Astronomy \& Astrophysics}\
  }\textbf {\bibinfo {volume} {574}},\ \bibinfo {pages} {17} (\bibinfo {year}
  {2015})}\BibitemShut {NoStop}%
\bibitem [{\citenamefont {Alam}\ \emph {et~al.}(2017)\citenamefont {Alam},
  \citenamefont {Ata}, \citenamefont {Bailey}, \citenamefont {Beutler},
  \citenamefont {Bizyaev} \emph {et~al.}}]{al}%
  \BibitemOpen
  \bibfield  {author} {\bibinfo {author} {\bibfnamefont {S.}~\bibnamefont
  {Alam}}, \bibinfo {author} {\bibfnamefont {M.}~\bibnamefont {Ata}}, \bibinfo
  {author} {\bibfnamefont {S.}~\bibnamefont {Bailey}}, \bibinfo {author}
  {\bibfnamefont {F.}~\bibnamefont {Beutler}}, \bibinfo {author} {\bibfnamefont
  {D.}~\bibnamefont {Bizyaev}}, \emph {et~al.},\ }\href@noop {} {\bibfield
  {journal} {\bibinfo  {journal} {Mon. Not. Roy. Astron. Soc.}\ }\textbf
  {\bibinfo {volume} {470}},\ \bibinfo {pages} {2617} (\bibinfo {year}
  {2017})}\BibitemShut {NoStop}%
\bibitem [{\citenamefont {Abbott}\ \emph {et~al.}(2019)\citenamefont {Abbott},
  \citenamefont {Allam}, \citenamefont {Andersen}, \citenamefont {Angus},
  \citenamefont {Asorey} \emph {et~al.}}]{tmc}%
  \BibitemOpen
  \bibfield  {author} {\bibinfo {author} {\bibfnamefont {T.~M.~C.}\
  \bibnamefont {Abbott}}, \bibinfo {author} {\bibfnamefont {S.}~\bibnamefont
  {Allam}}, \bibinfo {author} {\bibfnamefont {P.}~\bibnamefont {Andersen}},
  \bibinfo {author} {\bibfnamefont {C.}~\bibnamefont {Angus}}, \bibinfo
  {author} {\bibfnamefont {J.}~\bibnamefont {Asorey}}, \emph {et~al.},\
  }\href@noop {} {\bibfield  {journal} {\bibinfo  {journal} {The Astrophysical
  Journal Letters}\ }\textbf {\bibinfo {volume} {872}},\ \bibinfo {pages} {2}
  (\bibinfo {year} {2019})}\BibitemShut {NoStop}%
\bibitem [{\citenamefont {Alam}\ \emph {et~al.}(2015)\citenamefont {Alam},
  \citenamefont {Albareti}, \citenamefont {Prieto}, \citenamefont {Anders},
  \citenamefont {Anderson} \emph {et~al.}}]{fda}%
  \BibitemOpen
  \bibfield  {author} {\bibinfo {author} {\bibfnamefont {S.}~\bibnamefont
  {Alam}}, \bibinfo {author} {\bibfnamefont {F.~D.}\ \bibnamefont {Albareti}},
  \bibinfo {author} {\bibfnamefont {C.~A.}\ \bibnamefont {Prieto}}, \bibinfo
  {author} {\bibfnamefont {F.}~\bibnamefont {Anders}}, \bibinfo {author}
  {\bibfnamefont {S.~F.}\ \bibnamefont {Anderson}}, \emph {et~al.},\
  }\href@noop {} {\bibfield  {journal} {\bibinfo  {journal} {The Astrophysical
  Journal Supplement Series}\ }\textbf {\bibinfo {volume} {219}},\ \bibinfo
  {pages} {1} (\bibinfo {year} {2015})}\BibitemShut {NoStop}%
\bibitem [{\citenamefont {Goenner}(1984)}]{hf}%
  \BibitemOpen
  \bibfield  {author} {\bibinfo {author} {\bibfnamefont {H.~F.~M.}\
  \bibnamefont {Goenner}},\ }\href@noop {} {\bibfield  {journal} {\bibinfo
  {journal} {Foundations of Physics}\ }\textbf {\bibinfo {volume} {14}},\
  \bibinfo {pages} {9} (\bibinfo {year} {1984})}\BibitemShut {NoStop}%
\bibitem [{\citenamefont {Harko}\ \emph {et~al.}(2018)\citenamefont {Harko},
  \citenamefont {Koivisto}, \citenamefont {Lobo}, \citenamefont {Olmo},\ and\
  \citenamefont {D.R.Garcia}}]{th}%
  \BibitemOpen
  \bibfield  {author} {\bibinfo {author} {\bibfnamefont {T.}~\bibnamefont
  {Harko}}, \bibinfo {author} {\bibfnamefont {T.~S.}\ \bibnamefont {Koivisto}},
  \bibinfo {author} {\bibfnamefont {F.~S.}\ \bibnamefont {Lobo}}, \bibinfo
  {author} {\bibfnamefont {G.~J.}\ \bibnamefont {Olmo}},\ and\ \bibinfo
  {author} {\bibnamefont {D.R.Garcia}},\ }\href@noop {} {\bibfield  {journal}
  {\bibinfo  {journal} {Phys. Rev. D}\ }\textbf {\bibinfo {volume} {98}},\
  \bibinfo {pages} {084043} (\bibinfo {year} {2018})}\BibitemShut {NoStop}%
\bibitem [{\citenamefont {Camarena}\ and\ \citenamefont {Marra}(2020)}]{vm}%
  \BibitemOpen
  \bibfield  {author} {\bibinfo {author} {\bibfnamefont {D.}~\bibnamefont
  {Camarena}}\ and\ \bibinfo {author} {\bibfnamefont {V.}~\bibnamefont
  {Marra}},\ }\href@noop {} {\bibfield  {journal} {\bibinfo  {journal} {Phys.
  Rev. Research}\ }\textbf {\bibinfo {volume} {2}},\ \bibinfo {pages} {013028}
  (\bibinfo {year} {2020})}\BibitemShut {NoStop}%
\end{thebibliography}%
\end{document}